\documentclass[12pt,a4paper]{article}
\usepackage{amsmath}
\usepackage{amsfonts}
\usepackage{multicol}
\usepackage[dvips]{graphicx}
\usepackage{fixltx2e}

\newcommand{\danger}[1]{\textbf{#1}}

\input{epsf}             

\begin{document}

\title{\danger{Entropy in Spin Foam Models: The Statistical Calculation}}
\author{\centerline{\danger{J. Manuel Garc\'\i a-Islas \footnote{
e-mail: jmgislas@leibniz.iimas.unam.mx}}}  \\
Instituto de Investigaciones en Matem\'aticas Aplicadas y en Sistemas \\ 
Universidad Nacional Aut\'onoma de M\'exico, UNAM \\
A. Postal 20-726, 01000, M\'exico DF, M\'exico\\}

\maketitle

\begin{abstract}
Recently an idea for computing the entropy of black holes in the spin foam formalism has
been introduced. Particularly complete calculations for the three dimensional euclidean BTZ black hole 
were done. The whole calculation is based on observables living at the horizon of the black hole universe.
Departing from this idea of observables living at the horizon, we now go further and compute the
entropy of BTZ black hole in the spirit of statistical mechanics. We compare both calculations
and show that they are very interrelated and equally valid. This latter behaviour is certainly due
to the importance of the observables. 
\end{abstract}

\section{Introduction}

A way to compute the entropy of a black hole in the spin foam model description of quantum gravity has been introduced \cite{gi1}, \cite{gi2}. In particular complete calculations were
done for the case of the three dimensional euclidean BTZ black hole \cite{gi2}. 
It is also important to point out to approaches that go on the same spirit. They compute the 
geometrical entropy associated to microstates of spin networks; in \cite{lt},
the black hole entropy is studied in terms of quantum gravity and 
quantum information, that is in terms of entanglement of states, and in \cite{ma}, \cite{ma2}
the entropy of a black hole is studied in terms of quantum surface states.

Here in the same context of our previous studies and as a continuation of our research program 
we now compute the entropy of the BTZ black hole \cite{btz}
in the language of statistical mechanics. Departing form the observables idea
which was developed in our previous work we construct the statistical partition function 
of our black hole universe(system). 

We should of course compare both
calculations and hope they give the same or at least equivalent results. We indeed show that 
they give the same results up to a constant factor. This leads us to conclude that 
we may be in the correct direction when thinking of a way to study entropy
in the spin foam models.     

We divide this paper as follows.
In section 2 we briefly review the calculation of the entropy of 
euclidean BTZ black hole in the program of spin foam models of quantum gravity.

In section 3 we continue with our program for deriving a satisfactory way for computing
the entropy of black holes in the spin foam models of quantum gravity.  We derive
a computation in the spirit of statistical mechanics. Moreover, 
our idea is based
on the observables which live at the horizon and which were part of our
calculation in previous work.

Finally in section 4 we conclude with a discussion of our paper and other related ideas.   

\section{BTZ black hole entropy and spin foams}

We briefly review our idea for calculating the entropy of the euclidean BTZ black hole
in the context of spin foams. For a nice introduction to spin foam models based on quantum groups see
\cite{jb}.

The euclidean BTZ black hole has an hyperbolic type metric which after some identifications 
it is topologically a solid torus with its horizon at the core \cite{btz}.
For a deep study of the BTZ black hole see \cite{c}.
 
In the three dimensional case we have that
the
entropy of the BTZ black hole is given by

\begin{equation}
S \sim 2 L
\end{equation}
see \cite{c}.
After restoring factors $\hbar$ and $G$ it can be rewritten as

\begin{equation}
S \sim \frac{L}{4\hbar G}
\end{equation}
The horizon is the core of the torus.
We start by triangulating the Euclidean black
hole. We 
consider triangulations of the solid torus containing interior
edges, as we want the core of the torus(horizon) be formed by edges. 

\begin{figure}[h]
\begin{center}
\includegraphics[width=0.5\textwidth]{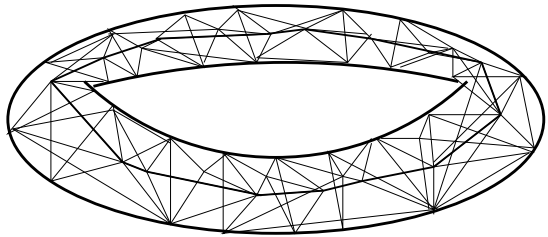}
\caption{Triangulation of the BTZ Euclidean black hole together with its horizon}
\end{center}
\end{figure}
The idea is based on the three dimensional topological spin foam models where the
irreducible representations of the quantum group $SU_{q}(2)$ play a role.
They are given by the finite set $\{ 0, \frac{1}{2}, 1, \frac{3}{2} \cdots \frac{r-2}{2} \}$ 
where $r \geq 3$.

A spin foam partition function of any triangulated three dimensional space-time $M$ 
is given by

\begin{equation}
Z(M)=  \sum_{S} \prod_{edges} dim_{q}(j) \prod_{tetrahadra} \{ 6j \}
\end{equation}
where the sum is carried over the set of all admissible states
$S$ and the
amplitude $\{ 6j \}$ is the $6j$ symbol associated to the six labels of
each tetrahedron. The quantum dimension is an amplitude associated to the edges.

The above partition function clearly can be applied to the BTZ black hole space-time
$M=T^2$. But now
consider the horizon of this black hole. We now think of the horizon as an observable
$\mathcal{O}$ in the following sense. Given the triangulation we consider the spin foam partition function with the difference that now we do not sum over the spins which label the horizon.

\begin{equation}
Z(T^2, \mathcal{O}(j_{1}, j_{2},...,j_{k})) = \sum_{S \mid
\mathcal{O}} \prod_{edges}dim_{q}(j) \prod_{tetrahedra} \{ 6j \}
\end{equation}
The spin foam approach is analogous to a Feynman path integral. In this case it is
the Feynman path integral of space-time. We therefore define
the expectation value or correlation function as

\begin{equation}
W(T^2, \mathcal{O})= \frac{Z(T^2, \mathcal{O})}{Z(T^2)}
\end{equation}
This is therefore a function of the labels of the horizon. 
It is shown in \cite{gi2} that for a particular triangulation of BTZ we have

\begin{equation}
W(T^2, \mathcal{O})= \prod_{m}
\frac{N_{i_m,j_{m},\widehat{j_m}}}{dim_{q}(\widehat{j_m})}
dim_{q}(i_m) dim_{q}(j_m)
\end{equation}
where $i_{1},j_{1} ,\cdots ,i_{n},j_{n}$ are spins labelling the horizon, and
$\widehat{j_1}, \cdots, \widehat{j_n}$ are spins which label edges that do not belong to the horizon,
however each triple $\{ i_{m}, j_{m}, \widehat{j_{m}} \}$ forms a triangle.    
Our particular triangulation has a horizon with an even number of edges. 

We propose that the entropy is given by the logarithm of formula $(5)$

\begin{equation}
S= \sum_{m=1}^{2n} \log(dim_{q}(j_m)) -
\sum_{k=1}^{n}\log(dim_{q}(\widehat{j_k)})
\end{equation}
It can be seen that the main contribution is given when the spins $\widehat{j_{m}}$ are zero.
This implies that each pair of the edges of the horizon are equal 
$i_{1}=j_{1}, i_{2}=j_{2}, \cdots  i_{n}=j_{n}$. 
The labels of the
edges of the horizon by spins $j$ are interpreted in the spin foam
model as giving a discrete length given by $j + \frac{1}{2}$.

The horizon is discrete and formed by edges with spins $j_1,\cdots
j_{2n}$.\footnote{We have just relabel edges $i_{1},j_{1} ,\cdots ,i_{n},j_{n}$ by 
 $j_{1}, j_{2} ,\cdots ,j_{2n}$} 

We have a constraint, since we want the sum of all the
discrete lengths of the horizon be $L$

\begin{equation}
\bigg(j_{1}+ \frac{1}{2} \bigg) + \bigg(j_{2}+ \frac{1}{2} \bigg) \cdots + 
\bigg(j_{2n}+ \frac{1}{2} \bigg) = L
\end{equation}
Observe that particularly when all of the spins at the horizon are 
equal we have

\begin{equation}
2nj+ \frac{2n}{2}= L
\end{equation}
which implies that 

\begin{equation}
n=\frac{L}{(2j+1)}
\end{equation}
Consider the case in which the number of spins we have goes to infinity, that is, when 
we go from the quantum group $SU_{q}(2)$ to the classical one $SU(2)$, such that
$dim_{q}(j) \rightarrow dim(j)= (2j+1)$

The entropy is then given by

\begin{equation}
S \sim L\ \frac{2\log(2j+1)}{(2j+1)}
\end{equation}
Finally it can be seen that the main contribution is given when $j=1$, that is,

\begin{equation}
S \sim 2L \ \frac{\log(3)}{3} 
\end{equation}
which is proportional to the length of the horizon and to equation $(1)$. 

The spin foam model procedure described in this section uses the quantum group 
$SU_{q}(2)$; at the end we took the limit $q \longrightarrow 1$. Our calculation
can be thought as having a regularisation procedure; it could be interesting to think if
a different regularisation procedure could be used as an alternative method to derive
the entropy of the black hole in a similar spirit as ours.

\section{Entropy in terms of the statistical partition function}

In this section, following our previous idea of the calculation of the entropy in the spin foam models of quantum gravity, we continue our proposal by deriving the entropy in a statistical spirit. 
In other words, we now consider the statistical partition function of our model.

The horizon is an observable in our picture. This means that 
the microstates live at the horizon. 
And we saw that when considering the main contribution to the entropy in the spin foam formalism
we should only care about what happens at the horizon, that is, it only matters the spins which label them.  

Since our approach is based on the spin foam models with cosmological constant\footnote{Positive cosmological constant in this case although the 
BTZ black hole has negative cosmological constant, we are not concern with this issue here,
but the idea is what matters for us at the moment} we only have a finite number of half integer spins
$\{ 0, \frac{1}{2}, 1, \frac{3}{2} \cdots \frac{r-2}{2} \}$ where $r \geq 3$. 

These spins are interpreted as giving a discrete length to any labelled edge
which as we saw contribute to the entropy of the black hole. Recall that a spin $j$ labelling
an edge is
interpreted as having a discrete length $j + \frac{1}{2}$. For convenience we consider the
length written as $\frac{2j+1}{2}$, and call the integer $\l_{j}=2j+1$. 

Our horizon is labelled, and must have length $L$. Therefore formula $(8)$ of section $3$
must be satisfied.
We rewrite it here and therefore have that 

\begin{equation}
n_{1} \frac{\l_{0}}{2} + n_{2} \frac{\l_{1/2}}{2} + \cdots +n_{r-1}\frac{\l_{(r-2)/2}}{2} = L
\end{equation}
where clearly $\l_{0}=1 , \l_{1/2}=2, \cdots \l_{(r-2)/2} =(r-1)$ and $n_{1}$, $n_{2}$, $\cdots$
$n_{r-1}$ are the number of edges labelled with spins $0$, $1/2$, $\cdots$, $(r-2)/2$ respectively. 
Interpret formula $(13)$ as in the case of the harmonic oscillator where now we have
energy given by length and occupation numbers given by the finite set of
$\{ \l_{j} \}$. 

We can consider a Boltzman
parameter $\beta$ in which the half term in the sum $(13)$ is absorbed, that is, $\beta= \frac{1}{2kT}$. 
We think of our model as an isolated system, similar to a gas made of an arbitrary 
number of photons which obeys a Bose-Einstein statistics.
In our case the analogy is to
consider our horizon to be formed by an arbitrary number of edges and each edge can  
be in any length state, that is, we can have any number of edges labelled with spin
$1/2$, some others labelled with spin $1$, and so on.

Therefore the statistical partition function of our model is given by

\begin{align}
Z &= \sum_{n_{1},n_{2} \cdots n_{r-1}} \exp [- \beta (\l_{0}n_{1} + \l_{1/2}n_{2}+ \cdots \l_{(r-2)/2}
n_{r-1} ] \nonumber \\
&=\sum_{n_{1},n_{2} \cdots n_{r-1}} \exp [- \beta (n_{1} + 2n_{2}+ \cdots (r-1)n_{r-1}) ] 
\end{align}
We want to consider the case of a black hole with very large length, that is, when the length 
goes to infinity.
Recall that our set of spins is finite, therefore we need to consider
a horizon with a very large number of edges which implies that  
$n_{1}, n_{2}, \cdots, n_{r-1}$ are unrestricted and can go to infinity.

The partition sum can be rewritten 

\begin{align}
Z &= \sum_{n_{1}=0}^{\infty} \exp[- \beta (n_{1})] \sum_{n_{2}=0}^{\infty} \exp[- \beta (2n_{2})] 
\cdots  \sum_{n_{r-1}=0}^{\infty} \exp[- \beta ((r-1)n_{r-1}] \nonumber \\ 
 &= \prod_{m =1}^{r-1} \sum_{n_{m =0}}^{\infty} \exp[- \beta (m n_{m})]
 = \prod_{m =1}^{r-1} \sum_{n_{m =0}}^{\infty} \exp(- \beta)^{m n_{m}}
\end{align} 
If we consider the complex variable $z=\exp(- \beta)$ in such a way that $|z|<1$ we have that 

\begin{equation}
\sum_{n_{m =0}}^{\infty} \exp(- \beta)^{m n_{m}} = \frac{1}{1-e^{- \beta m}} 
\end{equation}
and the partition function is now given by

\begin{equation}
Z=  \prod_{m =1}^{r-1} \frac{1}{1-e^{- \beta m}} 
\end{equation} 
Before we continue let us very briefly make the following mathematical observation. 

The partition function
written as in formula $(17)$ is a generating function
of the number of partitions of the positive integers $\mathbf{Z}^{+}$ in terms of the finite set 
$\{\l_{0}=1, \l_{1/2}=2,\cdots ,\l_{(r-2)/2}=(r-1) \}$.
We already know that
this finite set is related to the set of the discrete lengths associated  to the spins from which we are labelling the edges of our horizon.\footnote{$j+ \frac{1}{2}=\l_{j}/2$}
This gives a very nice connection to analytic number theory \cite{n}. We will come back to
this relation later on in the following section.

We can also state that analogous to the gas made of photons the Planck distribution can be derived
from formula $(16)$ and it is given by

\begin{equation}
\bar{n}_{m} = \frac{1}{e^{\beta m}-1} 
\end{equation} 
which gives information about the statistical equilibrium. That is, the most probable partition
corresponds to these numbers, given by the lowest spins of the finite set
$\{\l_{0}=1, \l_{1/2}=2,\cdots ,\l_{(r-2)/2}=(r-1) \}$.

The free energy is given by $F=-kT \ln Z$.  Substituting $\beta= \frac{1}{2kT}$ we have

\begin{equation}
F= -kT \ln \bigg[ \prod_{m =1}^{r-1} \frac{1}{1-e^{- \frac{m}{2kT}}} \bigg]
= -kT \sum_{m=1}^{r-1} \ln \bigg( \frac{1}{1-e^{- \frac{m}{2kT}}} \bigg)
\end{equation} 
which gives 

\begin{equation}
F= kT \sum_{m=1}^{r-1} \ln (1- e^{- \frac{m}{2kT}})
\end{equation} 
The entropy is then given by  

\begin{align}
S&= - \frac{\partial F}{\partial T}= - \bigg[ k \sum_{m=1}^{r-1} \ln (1- e^{- \frac{m}{2kT}}) 
- kT  \sum_{m=1}^{r-1} \frac{\frac{m}{2kT^{2}}e^{- \frac{m}{2kT}}}{(1- e^{- \frac{m}{2kT}})} \bigg]
\nonumber \\
&= \frac{1}{2T} \sum_{m=1}^{r-1} \frac{m e^{- \frac{m}{2kT}}}{(1- e^{- \frac{m}{2kT}})}
- k \sum_{m=1}^{r-1} \ln (1- e^{- \frac{m}{2kT}}) 
\end{align}
Now, technically the set of spins(equivalently the set of $\{ \l_{j}\}$) we have 
depend on the quantum group we choose $SU_{q}(2)$. We can choose to take the limit
when $q \rightarrow 1$, that is when going to
the classical group $SU(2)$; just as we did at the end of our calculation in section
$2$. 

Then the entropy can be considered to be given by

\begin{equation}
S= \frac{1}{2T} \sum_{m=1}^{\infty} \frac{m e^{- \frac{m}{2kT}}}{(1- e^{- \frac{m}{2kT}})}
- k \sum_{m=1}^{\infty} \ln (1- e^{- \frac{m}{2kT}}) 
\end{equation}
If we go further and approximate the sums by integrals we have

\begin{equation}
S= \frac{1}{2T} \int_{1}^{\infty} \frac{m e^{- \frac{m}{2kT}} \mathrm{d}m}{(1- e^{- \frac{m}{2kT}})}
- k \int_{1}^{\infty} \ln (1- e^{- \frac{m}{2kT}}) \mathrm{d}m
\end{equation}
The integrals can be computed for example with a mathematical program 
and then obtain that the entropy is given by

\begin{align}
S&= \frac{1}{2T} \bigg( \frac{1}{2} + \frac{4 \pi^{2} k^2 T^2}{3} - 2kT \ln (1- e^{\frac{1}{2kT}}) 
-4 k^{2} T^{2} \mathrm{Li}_{2}(e^{\frac{1}{2kT}})\bigg) \nonumber \\
&- k \bigg(- \frac{1}{4k T} - \frac{2 \pi^{2}k T }{3}+\ln (1- e^{\frac{1}{2kT}}) - \ln (1- e^{\frac{-1}{2kT}})  
+2kT \mathrm{Li}_{2}(e^{\frac{1}{2kT}})\bigg)
\end{align}
where the terms $\mathrm{Li}_{2}(e^{\frac{1}{2kT}})$ are called polylogarithmic integrals given by

\begin{equation}
\mathrm{Li}_{2}(e^{\frac{1}{2kT}})= \int_{0}^{e^{\frac{1}{2kT}}} \frac{\ln (1-x)}{x} \mathrm{d}x 
\end{equation}
Arranging terms, the entropy is given by

\begin{equation}
S= \frac{4 \pi^{2} k^2 T}{6} +\frac{1}{2T} - 2k \ln (1- e^{\frac{1}{2kT}}) + k \ln (1- e^{\frac{-1}{2kT}})
- 4k^{2}T \mathrm{Li}_{2}(e^{\frac{1}{2kT}})
\end{equation}
Our spin foam model description of black hole entropy is a first step towards the microscopic
description in terms of this quantum gravity direction. 
We are also considering  an Euclidean black hole and our model is in fact a toy model.
Let us go a bit further and suppose for a moment that the relation between temperature and length is fulfil.
For the BTZ black hole the temperature is given by $T= \frac{R}{2 \pi}$, where $R$ is the radius
of the horizon, see for instance \cite{c}. In the present case, our statistical partition function
(14), can be thought as a mathematical partition of the number $2L$, which therefore in this case 
leads us to consider the temperature to be given by $T= \frac{R}{\pi}$. 
With this value of the temperature the entropy(first term) is therefore given by 

\begin{equation}
S\sim  \frac{4 \pi r}{6}   k^{2}= 2L \ \frac{k^2}{6} 
\end{equation}
which is indeed proportional to the formula obtained in our previous work \cite{gi2},
given here by formula (12). They will be exactly equal if we have units in which we consider
$k^{2} = 2 \log (3)$.  Moreover, after restoring the constants $\hbar$ and $G$ it can be argued that
we are indeed obtaining the correct entropy of $\frac{L}{4 \hbar G}$ up to a constant factor.

\section{Discussion}

We have calculated the entropy of the euclidean three dimensional black hole in 
the spin foam formalism \cite{gi2}. We reviewed it in section 2 and  
extended the idea to the spin foam statistical calculation based on the observables.
The observables
played a significant role. The interpretation is that it only matters what happens at the horizon,
when considering the entropy. 

Let us mention some interesting ideas that come up here. How can all our calculations be extended
to an isolated four dimensional black hole universe? That is, develop a spin foam approach to
entropy of four dimensional black holes \cite{gi1}, as it is done in Loop Quantum Gravity
\cite{cr}, \cite{kk}, \cite{abck}, \cite{dl}, \cite{cpb}, \cite{abpbv}.
  
We may have naively chosen units such that $k^2= 2 \log (3)$ so that our formulae 
$(12)$ and $(26)$ match. What does it have to do with the fact that 
the spectrum of a quantum non rotating black hole such as Schwarzschild
is evenly spaced containing a factor $\log (3)$ as discussed in \cite{sh}, and later
in \cite{od}.

Another interesting fact is the relation of counting microstates which account for a fixed area
and partitions of integers in the field of number theory, principally analytic \cite{n}.
For instance recent developments about these number theory relations for the case of Loop Quantum Gravity
were introduced by \cite{abpbv} and continued in \cite{bv1}, \cite{bv2}.

Now particularly for this paper the relation is as follows. 
Suppose we want to count the number of undistinguishable microstates which account for a
fixed length horizon $L$. Then we would have to proceed as follows.
 
The statistical partition function 
written as formula $(17)$, is the generating function for the number of partitions
of the positive integers in terms of the finite set $\{\l_{0}=1, \l_{1/2}=2,\cdots ,\l_{(r-2)/2}=(r-1) \}$.

Such formula can be expanded as follows

\begin{equation}
Z=  \prod_{m =1}^{r-1} \frac{1}{1-e^{- \beta m}}
= \sum_{n=0}^{\infty} p_{r-1}(n) e^{- \beta n}  
\end{equation}
where $p_{r-1}$ is the number of partitions of the number $n$ in terms of integers not exceeding
$r-1$, that is in terms of our finite set $\{\l_{0}=1, \l_{1/2}=2,\cdots ,\l_{(r-2)/2}=(r-1) \}$.

If considering a large horizon length $L$, then counting the number of partitions of
such number $L$ gives us back $p_{r-1}(L)$. What we would be really counting with $p_{r-1}(L)$
is the number of undistinguishable microstates of the black hole. This means that
we would not be distinguishing between a given microstate which accounts for a fixed length
and a permutation of this microstate. 

In this case the number of microstates goes asymptotically as follows

\begin{equation}
p_{r-1}(L)= \frac{L^{r-2}}{(r-1)!(r-2)!}
\end{equation}
as can be seen in \cite{n}. 
Therefore if we want to calculate the entropy when the microstates are thought as undistinguishable,
it is just mainly given by the logarithm of formula $(28).$

\end{document}